\pdfoutput=1
\documentclass[reqno]{amsart}
\usepackage{breakcites}
\usepackage{preamble}
\usepackage{macros}
\usepackage{categories}
\usepackage{jon-tikz}

\def\paragraph{\subsubsection*}

\title{An inductive-recursive universe generic for small families}
\author{Daniel Gratzer}
\date{\today}

\begin{document}

\begin{abstract}
  We show that it is possible to construct a universe in all Grothendieck topoi with injective codes
  {\`a la} \textcite{pujet:2022} which is nonetheless generic for small families. As a trivial
  consequence, we show that \TTObs{} admits interpretations in Grothendieck topoi suitable for use
  as internal languages.
\end{abstract}

\maketitle
\tableofcontents

\begin{remark}
  We shall assume the Grothendieck universe axiom throughout this note to ensure a plentiful supply
  of Grothendieck universes.
\end{remark}

In recent work, \textcite{pujet:2022} have provided a comprehensive \emph{observational type theory}
complete with a hierarchy of universes and proven their theory enjoys decidable type-checking and a
number of other pleasant results. In order to ensure this, \emph{op. cit.} requires a
number of counter-intuitive properties of the universes. Specifically, they require that the
type-constructors on the universe are injective. In the case of dependent products, this means that
given a proof $e : \Prod{a : A_0} B_0\prn{a} \sim \Prod{a : A_1} B_1\prn{a}$, one can \emph{always}
produce a pair of proofs:
\begin{mathpar}
  e_0 : A_0 \sim A_1
  \and
  e_1 : \Prod{a : A_0} B_0\prn{a} \sim B_1\prn{\Cast\prn{A_0,A_1,e_0,a}}
\end{mathpar}
In other words, they require that $\Prod{-} -$ is injective.

Semantically, this is far from natural. Imagine, for instance, that $A_0 = A_1 = \ObjInit{}$, so
that both function types are equivalent to $\ObjTerm{}$, regardless of the choice of $B_0$ or
$B_1$. One can easily construct a model where these types are identified, so that we have no hope of
producing $e_1$ in such a model. In fact, this small example already shows that \TTObs{} cannot be
given the standard set-theoretic model wherein the universe is realized by a Grothendieck
universe. \textcite{pujet:2022}, however, have shown that \TTObs{} admits a model in setoids by
using an inductive-recursive construction to model the universe. We show that this approach is
easily generalized to give a model of \TTObs{} in arbitrary Grothendieck topoi and that a simple
modification to the standard IR universe ensures that \TTObs{} forms the basis for a workable
internal language in all of these settings.

\begin{remark}
  In a concession to brevity and convenience we modify \TTObs{} in several ways to better fit our
  techniques. Firstly, we regard the type theory as a signature in some logical framework
  (generalized algebraic theories, QIITs, representable map categories, LCCCs, or the like) thereby
  drop all discussions of coherence and partial interpretation functions featured prominently in
  \textcite{pujet:2022}.

  Secondly, we work with the universes as \emph{strict {\`a} la Tarski universes}. Without this
  change its inconceivable to have models of the theory in more complex categories where the
  distinction between objects and morphisms cannot be blurred away. For a user, however, the gap is
  substantially smaller than one might fear. We can always add a largest $\Uni[\omega]$ universe
  {\`a} la Tarski and systematically replace genuine types in a program with codes in this
  universe. Even this change is unnecessary however, as a type-directed elaboration procedure can
  easily paper over the mismatches.
\end{remark}

\begin{remark}
  We have occasion in this note to discuss both strong and weak Tarski universes. A strong Tarski
  universe is the standard notion: a type $\Uni$ and an explicitly decoding function $\Dec$ which
  commutes with a choice of codes in $\Uni$ for dependent products, sums, \etc{} A weak Tarski
  universe requires the same operations, but only satisfies the commutativity conditions up to
  isomorphism. The latter tends to more natural to obtain categorically, and some
  implementation-work has shown the notion to be workable in practice~\parencite{cooltt}.
\end{remark}

\section{Modeling \TTObs{} through induction-recursion}
\label{sec:ir}

One can model \TTObs{} in $\SET$ by interpreting the universe not by a Grothendieck universe, but
instead by an \emph{inductive-recursive} (IR) universe~\parencite{dybjer:2000}. For our purposes, we
will focus on \emph{small} induction~\parencite{hancock:2013}, where the eliminator is valued in a
universe of types smaller than the inductive definition. More verbosely, a small inductive-recursive
definition is a pair of some inductively defined family $A : \Uni[1]$ defined simultaneously with a
function $\Mor[r]{A}{\Uni[0]}$. Importantly, while induction-recursion generally has remarkable
proof-theoretic strength, small induction-recursion is a fairly innocuous reasoning principle and
can be encoded in extensional type theory with indexed inductive types.

Importantly, small induction-recursion is still sufficient to define IR universes in a type theory
with universes:
\begin{gather*}
  \begin{make-data}{V}{\Uni[1]}
    \Con{bool} : V
    \\
    \Con{unit} : V
    \\
    \Con{pi} : \prn{A : V} \to \prn{r\prn{A} \to V} \to V
    \\
    \Con{sg} : \prn{A : V} \to \prn{r\prn{A} \to V} \to V
  \end{make-data}
  \\[0.2cm]
  \begin{alignedat}{3}
    &r : V \to \Uni[0] &&
    \\
    &r\prn{\Con{bool}} = \mathbf{2}
    &
    \qquad
    &r\prn{\Con{unit}} = \mathbf{1}
    \\
    &r\prn{\Con{pi}\prn{A,B}} = \Prod{a : r\prn{A}} r\prn{B\prn{a}}
    &
    \qquad
    &r\prn{\Con{sg}\prn{A,B}} = \Sum{a : r\prn{A}} r\prn{B\prn{a}}
  \end{alignedat}
\end{gather*}

As an inductive type $V$ admits an induction principle which we can use to prove that
$\Con{pi}$ is injective, just as we can show that the successor is injective. Consequently,
$V$ provides exactly the basis we need to interpret \TTObs{} into $\SET$.

Summarizing, to interpret \TTObs{} into $\SET$ we start with some Grothendieck universe
$\mathcal{U}$ and we then use small IR within $\mathcal{U}$ to define a new set $V$ equipped with
injective codes for the type-constructors, and use this new set $V$ to interpret the universe of
\TTObs{}. In fact, because small induction-recursion lifts to Grothendieck topoi, this same approach
yields an interpretation of \TTObs{} into arbitrary topoi.

\subsection{The problem with $V$}

While this process yields a workable model of \TTObs{}, the model does not form the basis of a
good \emph{internal language}. In particular, because $V$ is defined by explicitly enumerating
the various constructors of the universe, $V$ lacks codes representing objects of the model laying
outside the image of the interpretation function. To pick a specific example, consider attempting
replaying the construction of \textcite{orton:2018} in \TTObs{}. We could not specialize the model
above to $\CSET$ to justify this development, because they require the universe to contain an
interval object and $V$ simply does not include such a constructor.

Of course, we could specialize the model in cubical sets further and explicitly include an interval
code to the definition of $V$. This is, however, hardly a satisfactory state of affairs! We do not
want a foundation for using \TTObs{} as an internal language that needs to be changed every time we
use a new aspect of our model.

We can quantify the problem more precisely by shifting our perspective on universes. While in type
theory a universe is a particular pair of a type and a family dependent over that type, in category
theory a universe is a collection of maps $\mathcal{S}$ stable under pullback and closed under
various operations~\parencite{streicher:2005}. One also requires a \emph{generic family} for such a
class---this is the categorical equivalent of what type theorists call a universe---but generic
families are not defined up to isomorphism and are not an invariant characteristic of universes.

We can phrase our issue with the IR universe $(V,r)$ somewhat more precisely by saying that it is
generic for a class $\mathcal{T}$ which lacks many important families in $\CSET$. In fact, an object
is classified by $\prn{V,r}$ only if it lies in the essential image of the unique functor logical
$\Mor{\II}{\CSET}$, where $\II$ is the initial elementary topos with a natural number object. This
is clearly an issue if we aim to use the universe to axiomatize types specific to $\CSET$ or indeed
any topos $\EE$.

\subsection{A plausible solution}
Of course, no matter how we interpret the universe some families in $\EE$ will lay outside
it. Indeed, for set-theoretical reasons we cannot hope to find a universe containing all families in
$\EE$, but we can hope for the next best alternative: a universe which contains all `small'
families.

The gold standard in this regard for Grothendieck topoi is to have a universe of all
\emph{relatively $\kappa$-compact} families~\parencite{shulman:2019}, where $\kappa$ is some
inaccessible cardinal. In fact, given a hierarchy of such universes for ever-increasing $\kappa$, we
can ensure that every family lies within some universe.\footnote{The Grothendieck universe axiom
  essentially stipulates this to be the case for $\SET$.} Helpfully, \textcite{streicher:2005} shows
that for all sufficiently large $\kappa$ this universe satisfies all the desirable
axioms. Crucially, \emph{op. cit.} shows that a generic family for the class of relatively
$\kappa$-compact morphisms exists in all Grothendieck topoi. Unfortunately, the supplied generic
family is based upon Grothendieck universes in $\SET$---precisely the generic family we just argued
cannot be used to interpret \TTObs{}.

Fortunately, generic families are not uniquely determined by a universe, and so we can hope for a
better one for the same class of morphisms:
\begin{conjecture}
  There is a generic families for relatively $\kappa$-compact morphisms equipped with injective
  codes for dependent products, sums, \etc{} in an arbitrary Grothendieck topos.
\end{conjecture}

\section{A generic family defined by small induction-recursion}
\label{sec:generic-family}

Fix some Grothendieck topos $\EE$ and a pair of inaccessible cardinals $\kappa_0 < \kappa_1$.
\textcite{streicher:2005} ensures that relatively $\kappa_i$-compact families organize into
universes $\mathcal{S}_{\kappa_{i}}$ in $\EE$ with generic families $\Mor[\El[i]]{\EL[i]}{\TY[i]}$
and inspection on the construction of the generic families reveals that $\Mor{\TY[0]}{\ObjTerm{}}$
is relatively $\kappa_1$-compact family. We will now construct a new generic family for
$\mathcal{S}_{\kappa_0}$ along with injective codes closing it under dependent products, sums,
\etc{}

To make this process a bit more fluid, we work in the \emph{internal language} of $\EE$. That is, we
work with an extensional type theory with a hierarchy of two weak universes \`{a} la Tarski
$\Uni[0] : \Uni[1]$.\footnote{We ignore strictness issues here, which can be rectified through any
  number of well-known constructions.} We will construct a universe
$\prn{V : \Uni[1],\Dec[V]{-} : V \to \Uni[0]}$ with the following operations:
\begin{itemize}
\item $\Mor[\Up]{\Uni[0]}{V}$ such that $\ArrId{} = \Dec[V] \circ \Up$.
\item $\FnCode : \Prod{a : V}\Prod{b : \Dec[V]{a} \to V} V$ such that $\FnCode$ is injective and
  $\Dec{\FnCode\prn{a,b}} = \Prod{x : \Dec{a}}{\Dec{b\prn{x}}}$.
\end{itemize}

\begin{remark}
  In fact, $V$ can trivially be extended to enjoy injective constructors similar to $\FnCode$ for
  dependent sums, booleans, equality types, \etc{} but we will focus on dependent products as a
  representative example.
\end{remark}

The first of these requirements ensures that $\prn{V,\Dec[V]}$ is generic for at least as many maps
as $\Uni[0]$ and the second gives the desired injective code for close $V$ under dependent
products. In fact, since $\Uni[0]$ is generic for a class of maps \emph{already} closed under
dependent products (though it does not necessarily witness this fact by an injective code) we can
conclude that $V$ is generic for precisely the same class as $\Uni[0]$.

Let us define $V$ by the following (small) inductive-recursive definition:
\begin{gather*}
  \begin{make-data}{V}{\Uni[1]}
    \Up : \Uni[0] \to V
    \\
    \FnCode : \prn{A : V} \to \prn{\Dec[V]{A} \to V} \to V
  \end{make-data}
  \\
  \Dec[V]{\Up\prn{A}} = A
  \qquad
  \Dec[V]{\FnCode\prn{A,B}} = \Prod{a : \Dec[V]{A}} \Dec[V]{B\prn{a}}
\end{gather*}

The two required functions are now just constructors of $V$ and both satisfy the required properties
simply by definition of $\Dec[V]$. Already from this simple construction we conclude the following:

\begin{theorem}
  \label{thm:one-universe}
  In an arbitrary Grothendieck topos $\EE$, there exists an interpretation of \TTObs{} with one weak
  universe {\`a} la Tarksi which sends the universe to a generic family for relatively
  $\kappa_0$-compact families.
\end{theorem}

\begin{remark}
  Notice here that we have obtained only a \emph{weak} universe, because $\FnCode\prn{A,B}$ decodes
  to a dependent product in $\Uni[0]$ which then must be lifted to $\Uni[1]$ to be regarded as a
  type in our model. Unfortunately, we have not assumed that code witnessing closure under dependent
  products in $\Uni[0]$ lifts to the equivalent code in $\Uni[1]$, and so do not obtain a model
  satisfying this equation. If we had assumed this however---and this requirement is satisfied by
  \eg{} the generic family supplied by \textcite{hofmann-streicher:1997}---we could correspondingly
  strengthen \cref{thm:one-universe}.
\end{remark}

\section{A strictly cumulative hierarchy}
\label{sec:hierarchy}

\Cref{thm:one-universe} is an excellent starting point, but we are interested in a hierarchy of such
universes. As before, we will show that we can `correct' a universe without injective codes to a one
with injective codes. We work in an arbitrary Grothendieck topos $\EE$. We work this time with a
hierarchy of inaccessible cardinals $\kappa_0 < \kappa_1 < \dots$. These induce a hierarchy of
universes $\Uni[0] : \Uni[1] : \dots$ in the extensional type theory of $\EE$, but unlike
\cref{sec:generic-family}, we will assume that we have constructed this hierarchy to be strictly
cumulative. This can be done in presheaf topoi using the construction of
\textcite{hofmann-streicher:1997}. In a general Grothendieck topos, one can use a more complex
construction of \textcite{shulman:elegant:2015}, which is discussed at length in forthcoming work by
Gratzer, Shulman, and Sterling.

We now proceed to inductively replace $\Uni[i]$ by $V_i$ such that $V_i$ is equipped with an
injective operation $\FnCode_i : \Prod{A : V_i} \prn{\Dec[V_i]{A} \to V_i} \to V_i$ and
$\prn{V_i,\Dec[V_i]}$ is generic for the same class of types as $\Uni[i]$, just as in
\cref{sec:generic-family}. We further ensure that there is an element $\UniCode[i] : V_j$ for all
$i < j$ such that $\Dec[V_j]{\UniCode[i]} = V_i$

Assume that $\prn{V_k : \Uni[k + 1], \Dec[V_k] : V_k \to \Uni[k]}$ has been defined for all $k <
i$. We define $V_i$ and $\Dec[V_i]$ as follows using small induction-recursion in $\Uni[k + 1]$:
\begin{gather*}
  \begin{make-data}{V_i}{\Uni[i + 1]}
    \Up : \Uni[i] \to V_i
    \\
    \UniCode[0], \cdots, \UniCode[i - 1] : V_i
    \\
    \FnCode : \prn{A : V_i} \to \prn{\Dec[V_i]{A} \to V_i} \to V_i
  \end{make-data}
  \\
  \Dec[V_i]{\Up\prn{A}} = A
  \qquad
  \Dec[V_i]{\UniCode[k]} = V_k
  \qquad
  \Dec[V_i]{\FnCode\prn{A,B}} = \Prod{a : \Dec[V_i]{A}} \Dec[V_i]{B\prn{a}}
\end{gather*}

It is plain that $V_i$ satisfies the required properties. As a final step, for each $i < j$ we
define a function $\Mor[\VLift]{V_i}{V_j}$ such that $\Dec[V_i]{A} = \Dec[V_j]{\VLift\prn{A}}$ and
so that $\VLift$ commutes with $\FnCode$ and $\UniCode[k]$. In fact, this specification fully
defines $\VLift$ and directly translates into a definition using the induction principle for $V_i$:
\begin{align*}
  &\VLift\prn{\Up\prn{A}} = \Up\prn{\Lift A}
  \\
  &\VLift\prn{\FnCode\prn{A,B}} = \FnCode\prn{\VLift\prn{A}, \VLift \circ B}
  \\
  &\VLift\prn{\UniCode[k]} = \UniCode[k]
\end{align*}

Inspection shows that $\VLift$ is functorial, and we thereby obtain the required strictly cumulative
hierarchy of universes.

\begin{theorem}
  \label{thm:universe-hierarchy}
  In an arbitrary Grothendieck topos $\EE$, there exists an interpretation of \TTObs{} with
  cumulative countable hierarchy of universes such that the $i$th universe is sent to a generic
  family for relatively $\kappa_i$-compact families.
\end{theorem}

In fact, we have really proven the following more general result:
\begin{theorem}
  A model of type theory with a cumulative hierarchy also supports a hierarchy with injective codes
  which remains generic for the same universes.
\end{theorem}

\section{Cumulativity from weak universes and induction-recursion}
\label{sec:real-ir}

Thus far our constructions have used only \emph{small} induction-recursion, so that the decoding
function associated with the inductive type targets a lower universe level. This restriction ensures
that the process can be decoded to indexed inductive types. If we assume, however, that we are
working in a model which supports true induction-recursion we can replicate
\cref{thm:universe-hierarchy} without assuming the input universe hierarchy is strictly cumulative.

We feel this construction is potentially interesting for constructivists; a constructively
acceptable version of the universes introduced by \textcite{shulman:elegant:2015} remains elusive,
and so a strictly cumulative hierarchy of universes in arbitrary Grothendieck topoi presently
requires choice. A priori, the same might not be true for induction-recursion and
\textcite{streicher:2005} has already shown that a hierarchy of universes which is merely
\emph{weakly} cumulative exists constructively. Accordingly, this construction offers an interesting
line of attack for a constructively acceptable hierarchy of universes in all Grothendieck topoi.

Let us fix a hierarchy of weak Tarski universes $\Uni[0] : \dots : \Uni[\omega]$. We proceed as
before and inductively replace $\Uni[i]$ by $V_i$ so that the latter equips the former with a strict
choice of codes. Unlike in \cref{sec:hierarchy}, we do not use small induction-recursion in
$\Uni[i + 1]$ in order to carry out this construction. Instead we use large IR in $\Uni[\omega]$
each time, and thereby avoid the need for a coherent choice of connectives in $\Uni[i]$.\footnote{In
  fact, we do not even require that $\Uni[i]$ be closed under \emph{any} connectives in this
  construction. We are freely closing the universe classified by $\Uni[i]$ with dependent products;
  if $\Uni[i]$ was already closed under dependent products this is an idempotent operation.}
\begin{gather*}
  \begin{make-data}{V_i}{\Uni[\omega]}
    \Up : \Uni[i] \to V_i
    \\
    \UniCode[0], \cdots, \UniCode[i - 1] : V_i
    \\
    \FnCode : \prn{A : V_i} \to \prn{\Dec[V_i]{A} \to V_i} \to V_i
  \end{make-data}
  \\
  \Dec[V_i]{\Up\prn{A}} = A
  \qquad
  \Dec[V_i]{\UniCode[k]} = V_k
  \qquad
  \Dec[V_i]{\FnCode\prn{A,B}} = \Prod{a : \Dec[V_i]{A}} \Dec[V_i]{B\prn{a}}
\end{gather*}

The lifting operation is define \emph{mutatis mutandis}.
\begin{theorem}
  A model with a weak hierarchy and induction-recursion can be extended to support a strict
  hierarchy generic for the same universes.
\end{theorem}

We emphasize the last point of this statement. It is well-known that large induction-recursion is
sufficient to define a cumulative hierarchy---this was the original example of IR---but we have
shown that our $\Up$ trick is sufficient to define a cumulative hierarchy which remains generic for
\eg{}, relatively $\kappa$-compact families. This point is unremarkable from within the type theory
itself, but crucial when using type theory as an internal language; it ensures that our universes
actually contain interesting families specific to a model.

\section*{Acknowledgments}
I am grateful for conversations with Carlo Angiuli and Jonathan Sterling.

\printbibliography

\end{document}